\documentclass[twocolumn,superscriptaddress,english,prl,aps]{revtex4-1}
\usepackage[utf8]{inputenc}
\usepackage{amsmath}
\usepackage{amssymb}
\usepackage{graphicx}
\usepackage{xspace}
\usepackage{mathtools}
\usepackage{physics}
\usepackage{bm}

\usepackage[backref=none,
bookmarksnumbered=true,
bookmarks=true,
bookmarksopen=true,
colorlinks=true,
citecolor=blue,
linkcolor=blue,
anchorcolor=green,
urlcolor=blue,unicode=false]{hyperref}

\begin{document}

\title{Resistivity tensor of vortex-lattice states in Josephson junction arrays}

\author{Alexander-Georg Penner}
\affiliation{\mbox{Dahlem Center for Complex Quantum Systems and Fachbereich Physik, Freie Universit\"{a}t Berlin, 14195 Berlin, Germany}}

\author{Karsten Flensberg}
\affiliation{\mbox{Center for Quantum Devices, Niels Bohr Institute, University of Copenhagen, DK-2100 Copenhagen, Denmark}}

\author{Leonid I.\ Glazman}
\affiliation{Department of Physics, Yale University, New Haven, Connecticut 06520, USA}

\author{Felix von Oppen}
\affiliation{\mbox{Dahlem Center for Complex Quantum Systems and Fachbereich Physik, Freie Universit\"{a}t Berlin, 14195 Berlin, Germany}}

\begin{abstract}
Two-dimensional Josephson junction arrays frustrated by a perpendicular magnetic field are predicted to form a cascade of distinct vortex lattice states. Here, we show that the 
resistivity tensor provides both structural and dynamical information on the vortex-lattice states and intervening phase transitions, which allows for experimental identification of these symmetry-breaking ground states. We illustrate our general approach by a microscopic theory of the resistivity tensor for a range of magnetic fields exhibiting a rich set of vortex lattices as well as transitions to liquid-crystalline vortex states. 
\end{abstract}

\maketitle

{\em Introduction.---}Josephson junction arrays display a fascinating variety of classical and quantum phases and phase transitions \cite{Ciria1998,Fazio2001}. In the absence of charging effects, the system undergoes a  temperature-driven Berezinskii-Kosterlitz-Thouless transition between a superconducting and a resistive phase. When the charging energy becomes large, quantum fluctuations  disorder the phase degree of freedom of the superconducting islands already at zero temperature, inducing a superconductor-insulator quantum phase transition. Particularly rich phase structure appears in a perpendicular magnetic field, even in the classical limit \cite{Pannetier1984,vanWees1987,Vu1993,Hallen1993}. The magnetic field introduces frustration into the effective description of the array in terms of an  $xy$-model. This   results in a cascade of different vortex-lattice ground states, which spontaneously break the underlying lattice symmetry of the array \cite{Teitel1983,Halsey1985}. While their equilibrium properties have been widely studied by numerical means  \cite{Teitel1983,Halsey1985,Kolahchi1991,Vallat1992,Straley1993,Franz1995,Gupta1998,Denniston1999,Lee2001,Lankhorst2018}, experimental studies have been lagging in identifying and probing these vortex-lattice states and little is known theoretically about their transport properties \cite{Falo1990,Otterlo1993}.  

Here, we show that the full resistivity tensor is a powerful tool to probe these vortex-lattice phases. Measurements of the longitudinal and transverse resistivity of Josephson junction arrays in a perpendicular magnetic field   reveal a dramatic sensitivity as a function of the magnetic field \cite{vanWees1987,Martinoli1993,Marcus2022}. This dependence reflects the sensitive dependence of the ground-state energy on the ratio $f$ of the magnetic flux $\phi$ per plaquette and the superconducting flux quantum $\phi_0=h/2e$. Our principal observations are twofold: First, the spontaneous breaking of the lattice symmetry  generally makes the resistivity tensor anisotropic, with the anisotropy encoding structural information on the vortex lattice. Second, phase transitions from vortex lattices to liquid-crystalline states \cite{Balents1995,Gupta1998} are signaled by the temperature dependence of the resistivity tensor. Importantly, full information is revealed only when measuring the entire resistivity tensor, which has not yet been done in experiments. 

Our work is motivated by recent advances in the nanofabrication of Josephson junction arrays based on semiconductor-superconductor hybrids \cite{Shabani2016}. These  arrays are exceptionally flexible in their lattice geometry and have  outstanding tunability of the junction properties \cite{Bottcher2022a,Bottcher2022b}. Work to date has focused on  the Berezinskii-Kosterlitz-Thouless transition \cite{Bottcher2022a} and vortex dynamics \cite{Bottcher2022b}. This generation of devices should readily allow for measurements of all components of the resistivity tensor, significantly advancing the experimental study of vortex lattices in Josephson junction arrays. 

\begin{figure}[b!]
    \centering
    \includegraphics[width=\columnwidth]{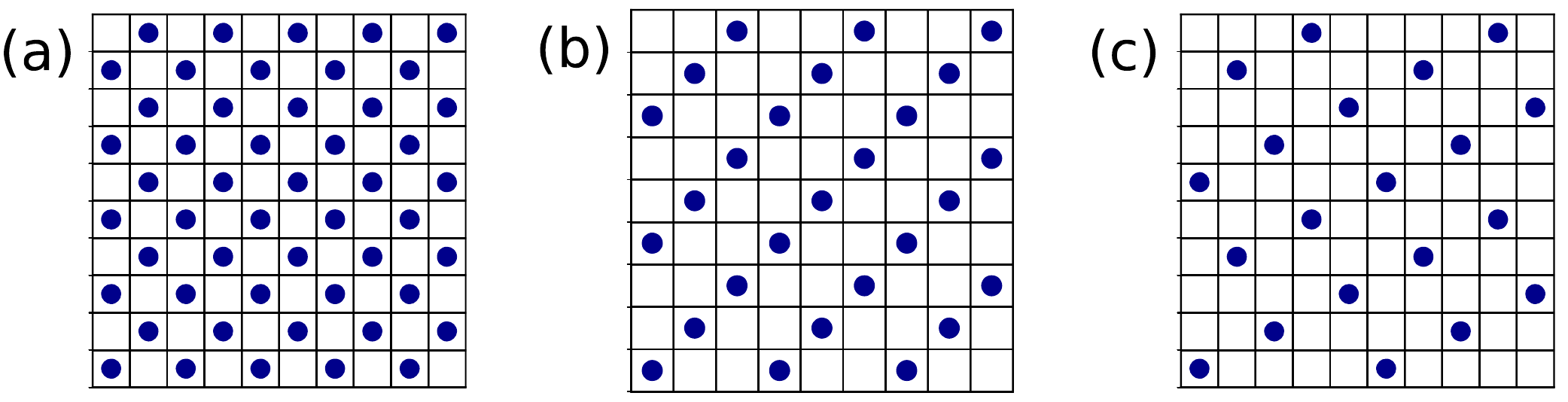}
    \caption{Vortex-lattice ground states for frustrations (a) $f=1/2$, (b) $f=1/3$, and (c) $f=1/5$.}   \label{fig1}
\end{figure}

{\em Model.---}We describe the Josephson junction array by viewing the superconducting islands as classical $xy$ spins. Neighboring spins are subject to an effective ferromagnetic interaction due to the Josephson coupling of strength $E_J$ between islands and the perpendicular magnetic field introduces frustration into the  $xy$ model, 
\begin{equation}
    H = - E_J \sum_{\langle ij\rangle} \cos (\varphi_i-\varphi_j +  2\pi a A_{ij}/\phi_0).
    \label{Eq:H}
\end{equation}
Here, $a$ is the lattice constant. We assume that the charging energies of the superconducting islands can be neglected, so that finding the equilibrium state of the array is a problem of classical statistical mechanics. The perpendicular magnetic field is accounted for by the vector potential $A_{ij}$, with the flux 
\begin{equation}
    \phi = f\phi_0 = \sum_{\mathrm{plaquette}} a A_{ij},
\end{equation}
per plaquette in units of the superconducting flux quantum $\phi_0=h/2e$ defining the frustration $f$. In the absence of the magnetic field, it is energetically favorable for the phases $\varphi_i$ of all superconducting islands to align. 

\begin{figure*}[t]       
\includegraphics[width=\textwidth]{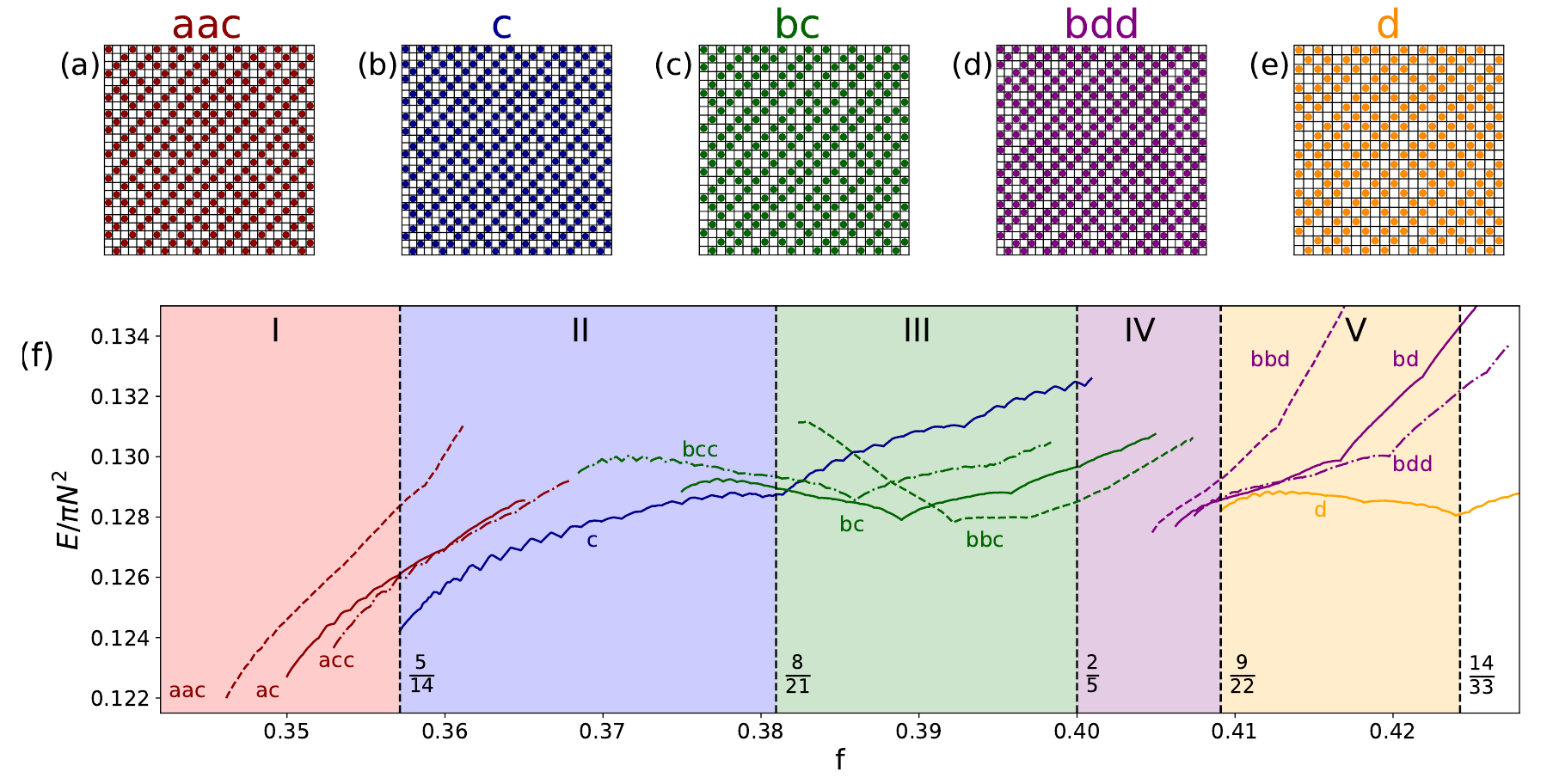}

    \caption{
(a-e) Vortex patterns for frustrations ${1}/{3}<\nu<{1}/{2}$. In this range, vortex lattices are built from a set of elementary units: a = 1/3 (filled diagonal enclosed by  empty diagonals on both sides), b = 2/5 (two filled diagonals enclosing an empty diagonal and enclosed by an empty diagonal on both sides), c = 5/14 (channel consisting of two 1/3 structures  enclosing a partially filled diagonal), d= 9/22 (thick channel consisting of two 2/5 structures enclosing a partially filled diagonal). Additional vortices can be accommodated into the channels by adding to the partially filled diagonals, changing the frustration $f$ by $1/N^2$. (f) Energy per lattice site of various vortex lattice structures as a function of $f$. The energy vs.\ $f$ is computed by gradually filling the channels with vortices. There are five color-coded regions I-V, separated by  dashed lines (with the corresponding $f$ indicated). Results are limited to short periods between channels. Note that this masks the transition at $f=2/5$, which occurs in the limit of large periods. } 
    \label{fig2}
\end{figure*}

The magnetic field introduces vortices, which have a natural inclination to order. The resulting  vortex lattice is a highly sensitive function of the magnetic field, reflecting the competition between the intrinsic tendency of vortices to order into a triangular structure and pinning by the lattice (which we take to be a square array). This is exemplified in Fig.\ \ref{fig1} for selected simple rational frustrations. For $f=1/2$, the vortices occupy every other site, breaking the underlying discrete translation symmetry of the array, but leaving its $C_4$ rotation symmetry intact [Fig.\ \ref{fig1}(a)]. For $f=1/3$, the vortices occupy plaquettes along every third diagonal, which also breaks the rotation symmetry down to $C_2$ [Fig.\ \ref{fig1}(b)]. At other frustrations such as $f=1/5$, the principal axes of the vortex lattice do not align along the diagonals of the square array [Fig.\ \ref{fig1}(c)]. 

{\em Resistivity tensor.---}For a square array of Josephson junctions, the resistivity will be isotropic as long as the ground state respects the underlying fourfold symmetry, e.g., at $f=\frac{1}{2}$. More generally, however, the formation of a symmetry-breaking vortex lattice will be signaled by an anisotropic resistivity tensor. The resistivity tensor is a sum of symmetric and antisymmetric contributions, 
\begin{equation}
    \rho = \left(\begin{array}{cc} \rho_{xx} & \rho^s_{xy} \\ \rho^s_{xy} & \rho_{yy} \end{array}\right) +
     \left(\begin{array}{cc} 0 & \rho_H \\ -\rho_H & 0 \end{array}\right).
\end{equation}
The symmetric contribution describes the dissipative resistivity and can be diagonalized with  eigenvalues $\rho_1$ and $\rho_2$. The antisymmetric contribution
is invariant under changes of the coordinate system and describes the Hall resistivity $\rho_H$ of the array. As a consequence of the Onsager relation $\rho_{xy}(B)=\rho_{yx}(-B)$, the two contributions $\rho_{xy}^s$ and $\rho_H$ to the transverse resistance $\rho_{xy}$ can be separated by their behavior under reversal of the perpendicular magnetic field $B$. While the Hall resistivity $\rho_H$ is odd under reversal of the magnetic field, $\rho_H(B) = - \rho_H(-B)$, the transverse resistivity $\rho^s_{xy}$ is even,
\begin{equation}
    \rho^s_{xy}(B) =  \rho^s_{xy}(-B).
\end{equation}
This is in line with the fact that the eigenvalues $\rho_j$ (with $j=1,2$) are even in magnetic field. 

It follows from this discussion that measurements of the full resistivity tensor reveal rich information on the vortex-lattice states. The eigenvalues $\rho_1$ and $\rho_2$ contain information on the density of mobile vortices and their mobility, while the principal axes reveal structural information about the vortex lattice. In particular, the observation of a transverse resistivity that is symmetric under reversal of the $B$ field is a direct indication of an anisotropic resistivity tensor and thus of spontaneous breaking of rotation symmetry by the vortex lattice. Interestingly, a $B$-symmetric transverse resistivity has been observed in experiments on vortex lattices \cite{vanWees1987,Marcus2022}.

We illustrate these general ideas by considering the resistivity tensor for a rich set of vortex lattice states appearing in the range of frustrations ${1}/{3}<f<{1}/{2}$. In this range, the vortex lattices are composed of a regular sequence of completely and partially filled as well as empty diagonals as illustrated in Fig.\ \ref{fig2} (a)-(e) \cite{Gupta1998,Lee2001}. Apart from the frustration $f$, we characterize the states by the fraction $p$ of partially filled diagonals as well as their filling $\nu$. The structure of the vortex lattices is simplest for ${5}/{14}<f<{8}/{21}$ [region II in Fig.\ \ref{fig2}(f)], where $p={1}/{7}$ and $f-p\nu={2}/{7}$ is the fraction of completely filled diagonals. However, our derivation below directly applies to any of the structures in Fig.\ \ref{fig2}. 

\begin{figure}
    \centering
    \hspace*{-.5cm}
    \includegraphics[width=.7\columnwidth]{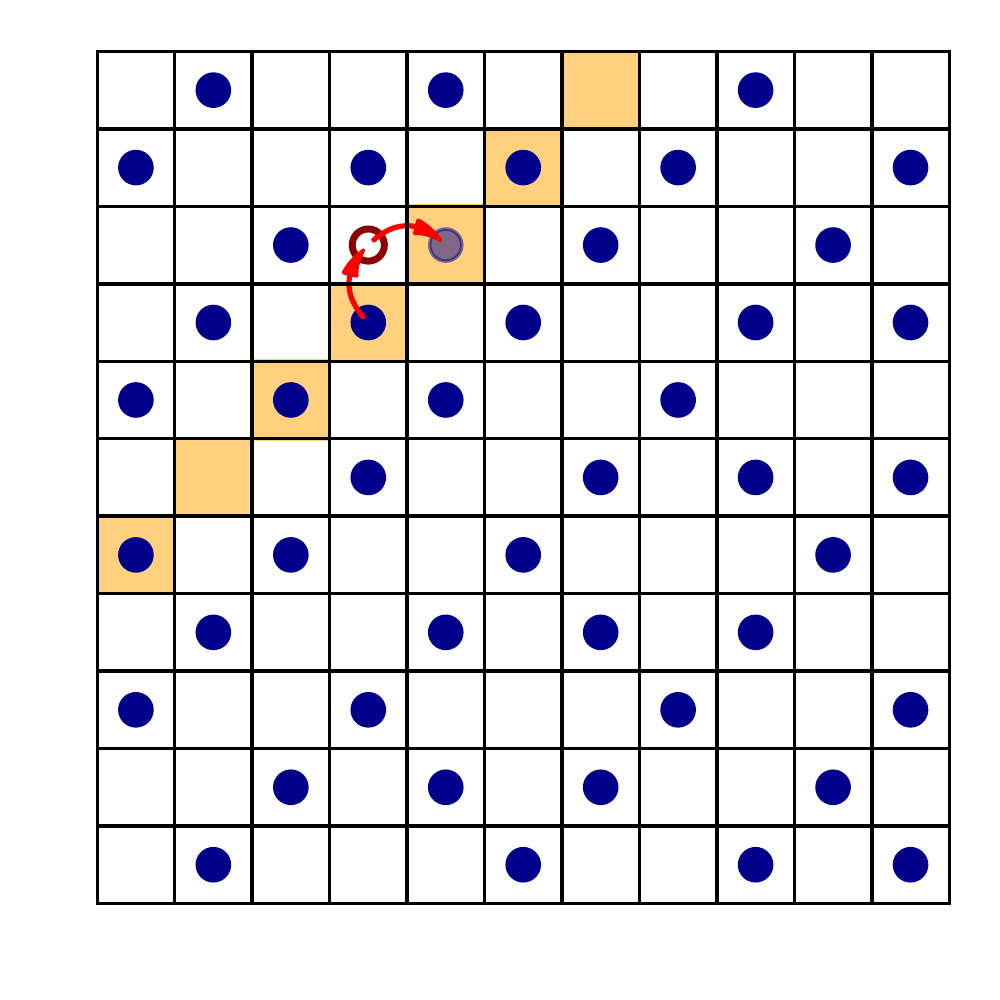}
    \caption{Hopping process of heavy domain wall in a c structure (blue region in Fig.\ \ref{fig2}). The initial vortex configuration (dark full circles) contains a heavy domain wall (see partially filled diagonal highlighted in orange). In the adjacent saddle-point configuration a vortex (empty circle) of the heavy domain wall hopped out of the partially filled diagonal. In the new adjacent minimum, the vortex (gray circle) returned to the partially filled diagonal and the heavy domain wall has moved by two sites. Red arrows indicate the hopping processes. We do not find minimum-energy configurations, in which vortices hop into the empty diagonals.}
    \label{fig3}
\end{figure}

The basic process governing the resistivity tensor is vortex hopping along the partially filled diagonals (Fig.\ \ref{fig3}). Stable vortex-lattice states have partially filled diagonals, which are at least half filled, $\nu>{1}/{2}$ (see \cite{SI}). In the low-temperature limit and for $\nu={1}/{2}$, the vortices organize into a regular array of alternating occupied and empty sites. Vortex motion becomes possible for $\nu>{1}/{2}$, for which there is a finite density of occupied nearest-neighbor sites (heavy domain walls). Starting with a minimal-energy configuration of Eq.\ (\ref{Eq:H}) with a heavy domain wall, we systematically search for adjacent saddle points (using the climbing-string method \cite{Ren2013,SI}). Subsequently, we search for new minima adjacent to the saddle point, in which the domain wall has moved by two sites. As illustrated in Fig.\ \ref{fig3}, we find that the basic process is a direct jump of the heavy domain wall along the diagonal. The saddle-point configuration has one vortex on a neighboring empty diagonal. We have not found a process with a final minimum-energy state, in which a vortex is located in an adjacent empty diagonal. From this calculation, we extract the height of the activation barrier for a hop of a heavy domain wall. For $\nu \simeq 1/2$, we find $E_B \simeq 0.83 E_J$. For larger $\nu$, the barrier increases slightly, e.g. for $f=13/35$ ($\nu = 3/5$), one finds an energy barrier of $E_B \simeq 1.01 E_J$. The thermally activated hopping rate of a heavy domain wall along the partially filled diagonal is $\Gamma\propto \exp\{-E_B/T\}$.

Since hopping of vortices is constrained to occur along the partially filled diagonals, the resistivity tensor will be strongly anisotropic, with principal axes aligned along and perpendicular to the partially filled diagonals. In the absence of vortex hopping out of the partially filled diagonals, there is no voltage drop along the diagonals and the resistivity eigenvalue $\rho_2$ for currents applied along the diagonals vanishes, $\rho_2=0$. In contrast, current applied perpendicular to the diagonals of the vortex lattice induces vortex motion along the  diagonals. This generates a voltage drop along the current direction, resulting in a nonzero resistivity eigenvalue $\rho_1$. 

To compute the nonzero eigenvalue $\rho_1$ of the resistivity tensor, we consider an applied current $j_c$ flowing perpendicular to the diagonals of the vortex lattice. Provided that current flow is uniform, this induces potential drops for vortices on neighboring sites in the horizontal ($\hat x$) and vertical directions ($\hat y$) direction, which are equal to \cite{Halperin1979}
\begin{equation}
    \Delta U = \frac{h}{2e} \frac{1}{\sqrt{2}}j_c a.
    \label{DeltaU}
\end{equation}
Equation (\ref{DeltaU}) can be understood on the basis of the potential $U(\varphi)=- E_J \cos\varphi - \frac{\hbar}{2e} i_b \varphi$ of a single junction with Josephson energy $E_J$, phase difference $\varphi$, and bias current $i_b$. A vortex hopping between two neighboring plaquettes leads to a phase slip of the intermediate junction by $2\pi$, which changes the potential by $\frac{h}{2e}i_b$. Alternatively, Eq.\ (\ref{DeltaU}) can be viewed as a manifestation of the Magnus force that the current exerts on the vortices.

For $\nu>{1}/{2}$, there are two heavy domain walls for each additional vortex. Thus, the areal density of domain walls is equal to $2p(\nu-\frac{1}{2})\frac{1}{a^2}$. When $\nu$ is not too much larger than ${1}/{2}$, the domain walls are dilute and the probability of two directly adjacent domain walls can be neglected. Accounting for the potential drop of $2\Delta U$ between initial and final state of the vortex hop in linear response, a vortex in a heavy domain wall has an effective hopping rate of $ (2\Delta U/T)\Gamma$. As a result, we find an areal vortex current density of 
\begin{equation}
    j_v = 2p(\nu-\frac{1}{2})\frac{\sqrt{2}}{a} \Gamma \frac{2\Delta U}{T} 
    \label{eq:jv}
\end{equation}
along the diagonals of the vortex lattice. We finally relate the vortex current density to the voltage drop in the perpendicular direction \cite{Halperin1979}, 
\begin{equation}
    E = \frac{h}{2e}j_v.
    \label{eq:E}
\end{equation}
The electric field $E$ points in the direction perpendicular to the vortex motion and thus perpendicular to the partially filled diagonal. Combining Eqs.\ (\ref{DeltaU}), (\ref{eq:jv}), and (\ref{eq:E}), we find the nonzero resistivity eigenvalue 
\begin{equation}
    \rho_1 = 4p(\nu-\frac{1}{2}) \left(\frac{h}{2e}\right)^2\frac{\Gamma}{T}
\end{equation}
in the limit of dilute domain walls. 

While the partially filled diagonals order into a regular pattern at low temperatures, numerical results indicate that these ordered arrays melt prior to the melting transition of the entire 2D vortex lattice \cite{Gupta1998}. If the interaction between vortices is sufficiently short ranged due to screening, the coupling between diagonals is weak for small $p$. Without coupling, the partially filled diagonals are expected to melt at any nonzero temperature by the Landau-Peierls argument. At weak coupling (small $p$), ordered states of the partially filled diagonals melt far below the melting temperature of the entire vortex lattice. In the intermediate regime between these melting transitions, the vortex state is akin to a smectic liquid crystal. 

Well above the melting transition, we can assume that the vortex occupations become uncorrelated along the partially filled diagonals. Vortex motion along the  diagonals can then be characterized by a diffusion constant $D=2a^2/\tau$ and the vortex current becomes  
\begin{equation}
    j_v = p\nu(1-\nu) \frac{1}{a^2}\frac{\sqrt{2}a}{\tau} \frac{2\Delta U}{T}, 
\end{equation}
so that 
\begin{equation}
    \rho_1 = 2p\nu(1-\nu)  \left(\frac{h}{2e}\right)^2\frac{1}{T\tau}.
\end{equation}
Unlike in the low-temperature phase, the hopping rate $1/\tau$ is no longer activated, so that the melting transition is signaled by a substantial increase in resistivity. 

We now consider the resistivity tensor for currents applied  along the lattice directions of the underlying square array. Rotating the resistivity tensor, we find 
\begin{equation}
    \rho = \left(\begin{array}{cc}
    \rho_1/2 & \pm\rho_1/2 \\
    \pm\rho_1/2 & \rho_1/2
    \end{array}\right). 
\end{equation}
The sign of the transverse resistivities depends on the direction of the diagonal vortex structure. We also use that  $\rho_H=0$ within our rate-equation theory of vortex dynamics. Due to the diagonal structure of the vortex lattice, the diagonal resistivities $\rho_{xx}$ and $\rho_{yy}$ are identical. The formation of the vortex lattice is still signaled by the nonzero transverse resistivity. Since vortex motion is constrained to be along the diagonal, a current applied, say, along the $x$-axis induces equal voltage drops along the $x$ and $y$-directions. Thus, $\rho_{xx}$ and $\rho^s_{xy}$ are equal in magnitude, a feature that directly correlates with the diagonal structure of the vortex lattice.  

{\em Magnetic-field dependence.---} There will also be substantial and characteristic variations with magnetic field. This dependence emerges from the intricate sequence of vortex lattices for frustrations between $f=1/3$ and $f=1/2$ \cite{Gupta1998,Lee2001}. As shown in Fig.\ \ref{fig2}, the vortex lattices are built from  elementary units. There are units with only completely filled and empty diagonals corresponding to the basic units of the $f=1/3$ and $f=2/5$ vortex lattices. In addition, there are units referred to as channels, which  contain the partially filled diagonals (see Fig.\ \ref{fig2} and its caption for a detailed description). 

These structures were identified by Monte Carlo simulations in Refs.\ \cite{Gupta1998,Lee2001}. We also find such structures using a recently proposed annealing algorithm \cite{Lankhorst2018,SI}, which computes the full phase configuration. We can further confirm and extend these results using a vortex representation of the $xy$ model \cite{Jose1977,teitel2013two,SI}. Figure \ref{fig2}(f) compares the energy of various structures, computed numerically as a function of $f$ by adding vortices one at a time to the partially-filled diagonals. For a given regular pattern of building blocks, we use a Metropolis algorithm to optimize the vortex arrangement along and between partially filled diagonals. Our results essentially reproduce the conclusions of Ref.\ \cite{Lee2001}, but indicate the existence of an additional regime originating from the  previously unnoticed phase transition at $f=9/22$. 

The structure of the vortex lattice remains unchanged over a substantial range in magnetic field in the regions II (blue) and V (yellow). Here, $p$ remains fixed and  the resistivity depends smoothly on magnetic field. In contrast, the vortex-lattice structure, and hence $p$ as well as the resistivity tensor, depend sensitively on magnetic field in regions I (red), III (green), and IV (purple). Note that in these regions, there are more vortex-lattice states than shown in Fig.\ \ref{fig2}(f), which includes only short-sequence lattices. 

Our vortex-lattice simulations use periodic boundary conditions. In real samples, boundary effects as well as disorder may lead to domains of different orientations of the vortex lattice. One expects that this reduces the anisotropy of the resistivity and hence the magnitude of $\rho_{xy}^s$ relative to $\rho_{xx}$ and $\rho_{yy}$. The reduction would be a direct measure of the degree of domain formation of the vortex lattice. 

{\em Conclusions.---}We have shown that measurements of the full resistivity tensor are a powerful tool to identify and probe the rich set of symmetry-breaking vortex-lattice states as a function of magnetic field, providing both structural and dynamical information. This includes the liquid-crystalline vortex states above the melting transition of the partially filled diagonals, which have the remarkable property of being resistive in one direction and superconducting in the other. Our general approach carries over to other (nematic) electronic states which break the underlying rotational symmetry (for a very recent discussion, see Ref.\ \cite{Sau2023}). 

We have focused on vortex configurations aligned along the diagonal, which cover a wide range of frustrations. In these states,  the vortex dynamics is particular transparent. However, one in principle expects anisotropic  resistivity tensors also for other symmetry-breaking vortex-lattice states. This includes states such as the $f={1}/{5}$ state [see Fig.\ \ref{fig1}(c)], whose principal axes are rotated by other angles, so that $\rho_{xx}$ and $\rho_{yy}$ are different from each other. In this case, measurements of the full resistivity tensor may also contribute to elucidating the underlying vortex dynamics. 

Our discussion neglected capacitive effects and was limited to classical modeling of the Josephson junction arrays. While our qualitative conclusions are expected to persist, it is an interesting question for future work to relax these assumptions and account for quantum fluctuations, which may, e.g., induce a nonzero Hall resistivity $\rho_H$.

\begin{acknowledgments}
We thank C.M.\ Marcus for discussions stimulating the present work. Financial support was provided by Deutsche Forschungsgemeinschaft through CRC 183, the Einstein Research Unit on Quantum Devices, the Danish National Research Foundation, the Danish Council for Independent Research $\vert$ Natural Sciences, the European Research Council (ERC) under the European Union’s Horizon 2020 research and innovation program under grant agreement No. 856526, the National Science Foundation through Grant No.~DMR-2002275, and the Office of Naval Research (ONR) under award number N00014-22-1-2764. L.I.G. thanks Freie Universit\"{a}t Berlin for hosting him as a CRC 183 Mercator fellow.  
\end{acknowledgments}


%

\onecolumngrid

\clearpage

\setcounter{figure}{0}
\setcounter{section}{0}
\setcounter{equation}{0}
\renewcommand{\theequation}{S\arabic{equation}}
\renewcommand{\thefigure}{S\arabic{figure}}

\onecolumngrid


\section*{Supplementary Information}

\section{Vortex representation}

The Josephson junction array described by Eq.~(\ref{Eq:H}) can alternatively be viewed in a dual vortex picture. Performing a Villain transformation \cite{Jose1977,teitel2013two}, one obtains the vortex Hamiltonian
\begin{equation}
    H = \frac{\pi E_J }{2} \sum_{ij} (m_i-f)G(\mathbf{r}_i-\mathbf{r}_j)(m_j-f),
    \label{vortex Hamiltonian}
\end{equation}
where $i$ and $j$ run over the $N^2$ lattice sites located at $\mathbf{r}_i$ and
\begin{equation}
    G(\mathbf{r}) = \frac{\pi}{N^2}\sum_{\mathbf{k} \neq 0} \frac{e^{i{\mathbf{k}\cdot \mathbf{r}}}-1}{2- \cos(k_x x) - \cos(k_y y)}.
\end{equation}
The components of $\mathbf{k}$ are quantized in multiples of $\frac{2\pi}{N}$. We use the vortex representation for the calculations underlying Fig.\ \ref{fig2}.

For a given combination of elementary building blocks of striped vortex lattices, e.g., aac, we compute the energy via Eq.~(\ref{vortex Hamiltonian}) for $\nu = 1/2$. We then increase the magnetic field in units of $1/N^2$ by adding vortices to the channels. We minimize the energy of a given lattice structure by means of a metropolis algorithm. We also use this algorithm to decide to which channel the additional vortex is added. Finally, we allow the partially filled diagonals to shift against each other, which significantly decreases the energies of the c and d structures. We use system sizes $N \sim 100$, chosen in  such a way that they are commensurate with the structure (i.e., for c structures, we choose system sizes which are a multiple of 14, for aac, $N$ is chosen as a multiple of 26, etc.) 
We find that the energies do not exhibit significant $N$ dependence if $N$ is increased further.

We comment on  trends evident in Fig.\ 2(f) of the main text. In the region $\frac{1}{3}<f<\frac{5}{14}$ (red), the building blocks of type a have a natural frustration $f=\frac{1}{3}$, while the channels c are stable for frustrations larger than $f=\frac{5}{14}$ corresponding to half-filled channels. This implies that  alternations of a and c are stable beyond minimal frustrations, which grow with the fraction of c building blocks. The energy of a particular structure grows rapidly with $f$, since both the building blocks of type a become energetically less favorable (they are the natural building blocks at $f=\frac{1}{3}$) and the channels c become filled beyond half filling, which adds domain walls. We note that Fig.\ 2(a) contains only short-period structures. Other, longer-period structures are likely ground states in some region of frustrations $f$, although they are not included in the figure.  

\begin{figure*}[b!]
    \centering
    \includegraphics[width=.8\textwidth]{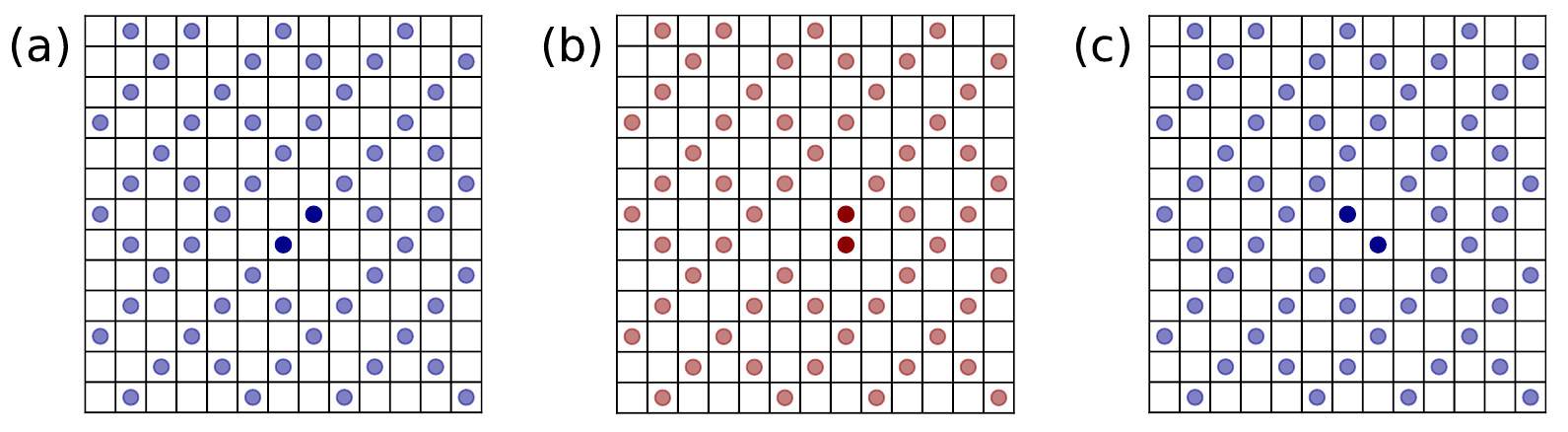}
    \caption{ Instability of channels for occupations $\nu < 1/2$ of the partially filled diagonal. (a) Metastable structure with light domain wall. Light domain walls necessarily exist for $\nu<1/2$. (b) Adjacent saddle-point configuration with a vortex displaced out of a neighboring fully occupied diagonal. (c) Lower-energy minimum reached from the saddle point, in which two vortices have moved out of full diagonal.}
\label{fig:wallflip}
\end{figure*}

It is, at least in principle, possible  that in some restricted regions of $f$, the true ground state is not of this striped form, but has a more complicated 2d structure. For instance, partially filled diagonals which are less than half filled destabilize the walls of the channels (see Fig.\ \ref{fig:wallflip} for corresponding numerical results). It is conceivable that this may destabilize the stripe-like pattern into a more complicated 2d configuration just before the onset of one of the structures. 

In the region $\frac{5}{14}<f<\frac{8}{21}$ (blue), the lowest energy structure consists of channels c only, with the partially filled diagonal filling up from being half filled ($f=\frac{5}{14}$) to being two-third filled ($f=\frac{8}{21}$). Thus, one expects that the energy of the state increases as $f$ grows within this regime, as indeed observed. Note that beyond $f=\frac{8}{21}$, the partially filled diagonals necessarily contain instances of three consecutive occupied sites. This induces a noticable cusp in the energy of the c configuration vs.\ frustration.   

In the region $\frac{8}{21}<f<\frac{2}{5}$ (green), we find structures that combine channels c with the basic unit b of the $f=\frac{2}{5}$ ground state. As $f$ increases within this region, one thus expects that the b blocks become energetically more favorable, while the c blocks become energetically more costly as the partially filled diagonal increase in occupancy. This is consistent with the weaker $f$ dependence of the configurations combining b and c blocks. 

We also note that the bc structure can be competitive with a pure c structure near $f=\frac{8}{21}$, since the filling of the channel is much lower in bc than in c at these frustrations. Similarly, the bcc structure can have a higher energy than the bc structure, since the channels have a larger occupation in bcc than in bc. Eventually, this trend would have to reverse however, as structures of the form bc$\ldots$c should become close in energy to a pure c configuration, as the number of c building blocks increases. 

In the purple region, one expects a transition from  channels c to thick channels of type d interspersing b-type building blocks. This transition will presumably happen in the limit of rare interspersing (thick) channels, so that it is masked in Fig.\ 2, which is limited to short periods. 

In the region $\frac{9}{22}<f<\frac{14}{33}$ (orange; not previously identified), we find that the ground state is built from thick channels d, with the partially filled diagonal again varying from half to two-third filling. Compared to the channels c, we observe that the energy of the structure is more weakly dependent on the frustration $f$.

\begin{figure*}[b!]
    \centering
    \includegraphics[width=0.8\textwidth]{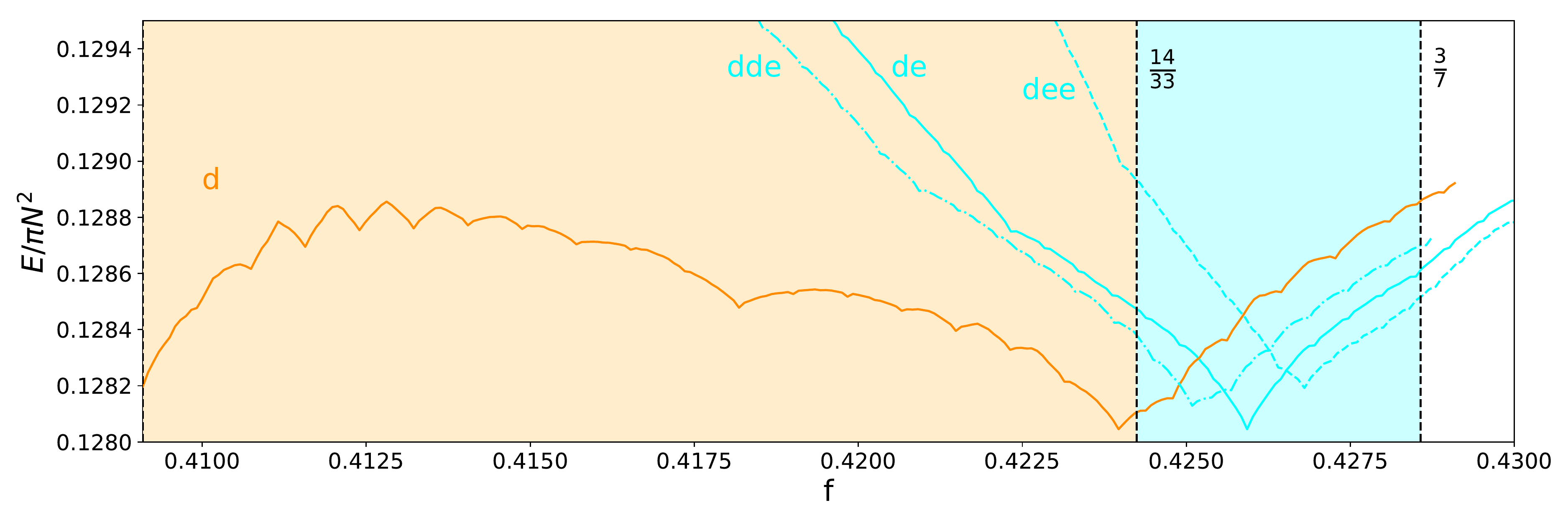}
    \caption{Energy per  lattice site of various vortex-lattice structures as a function of $f$. This figure extends Fig.\ 2(f) beyond $f=14/33$ up to $f=3/7$. In this region, the ground states likely consist of combinations of thick channels (building block $d$) and building block $e$ (three filled diagonals alternating with four empty  diagonals, corresponding to the elementary unit of the $f=\frac{3}{7}$ structure).}
    \label{groundstatebeyond}
\end{figure*}

Finally, we remark that in the white region in Fig.\ 2(f) beyond $f=\frac{14}{33}$, it is conceivable that the ground states are constructed from $f=\frac{3}{7}$ building blocks, interspersed with thick channels d. Corresponding energies are shown in Fig.\ \ref{groundstatebeyond}. We note that the $f=\frac{4}{9}$ stripe structure is no longer a stable ground state, but preempted by vacancy crystals on top of the $f=\frac{1}{2}$ state. This will eventually preclude additional regimes of stripe structures. 

\section{Phase configurations}

To search for minima and saddle points of the phase configurations, we start from  Eq.~(\ref{Eq:H}). We can search for the ground-state configuration using the algorithm proposed in Ref.~\cite{Lankhorst2018}. The algorithm relies on describing the junctions within the RCSJ model and solving the set of Kirchhoff's rules at each lattice site of the array. This results in $N^2$ coupled linear differential equations. Temperature is incorporated by  Langevin currents due to Johnson-Nyquist noise. Starting from a random initial configuration of the phases, one computes the time evolution of the system while temperature is gradually decreased to zero. We use periodic boundary conditions, which requires $N$ to be a multiple of $q$ for a frustration of $f=p/q$. 

In the range $f=1/3$ to $f=1/2$, a direct application of this algorithm \cite{Lankhorst2018} tends to result in configurations with domain walls. We avoid this by applying a slightly larger magnetic field along diagonals, where the expected vortex lattice has nonzero vortex occupation (and compensate on empty diagonals to leave the overall magnetic flux fixed). Following the annealing process, we revert to a uniform magnetic field and adjust the phase configuration using a steepest-decent algorithm. We find no change of the vortex configuration in this process.  

We use the following parameters in our simulations: Josephson coupling $E_J = 1$, normal resistance $R=1$, $n = 10^5$ time steps, step size $\Delta \tau = 0.5$, initial temperature $T_0 = 0.2$. For the first $2\times 10^4$ time steps, temperature is held constant after which it is decreased linearly to zero. For the stripe phases, we enhance the frustration on the expected fully occupied diagonals by $\Delta f= 0.1$ and decrease the frustration on one neighboring diagonal by the same amount. 

To identify the relevant hopping processes of vortices, we proceed as follows. We apply the climbing-string method described in the next section to the phase model in Eq.~(\ref{Eq:H}) to identify the adjacent saddle point. Subsequently, we follow the steepest descent to find the final-state minima. We also confirm the resulting activiation barriers using the vortex model. Note, however, that the vortex model describes a discrete set of vortex configurations. Thus, it cannot describe the entire activation path, but only initial, final, and saddle configurations. 

\section{Climbing-string method}

We compute the energy barrier between two minima of the vortex lattice using the climbing-string method, see Ref.~\cite{Ren2013}. The idea behind this saddle-point search is to start with a phase configuration close to a minimum and to reverse the force pushing the phases into the minimum. This converges to an adjacent saddle point in a controlled manner. We  briefly summarize the essential steps and detail the numerical values of parameters used in the simulations. 

\textit{Step 1:} We denote the minimum of the vortex lattice by the vector $\boldsymbol{\Phi}_0 = (\phi_1, \phi_2, ..., \phi_{N^2})$. We start with an initial  random configuration $\boldsymbol{\Phi}_M$ close to $\boldsymbol{\Phi}_0$. We linearly interpolate between these two configurations to obtain $M-1$ images $\boldsymbol{\Phi}_i$ with $i = 1, ..., M-1$, which form a string with the end points $\boldsymbol{\Phi}_0$ and $\boldsymbol{\Phi}_M$. We now keep $\boldsymbol{\Phi}_0$ fixed and evolve the images according to the equations of motion
\begin{align}
\dot{\boldsymbol{\Phi}}_i &= -\boldsymbol{\nabla} H, \ \ \ 0 < i < M \\
\dot{\boldsymbol{\Phi}}_M &= -\boldsymbol{\nabla} H + \theta (\boldsymbol{\nabla} H \cdot \boldsymbol{\tau})\boldsymbol{\tau}.
\end{align}
with $\theta > 1$ and the approximate tangent vector $\boldsymbol{\tau}$ given by 
\begin{equation}
\boldsymbol{\tau} = \frac{\boldsymbol{\Phi}_M - \boldsymbol{\Phi}_{M-1}}{|\boldsymbol{\Phi}_M - \boldsymbol{\Phi}_{M-1}|}.
\end{equation}
The endpoint $\boldsymbol{\Phi}_M$ is thus climbing up the basin of the minimum $\boldsymbol{\Phi}_0$ in the tangent direction of the string, while following steepest decent dynamics in the perpendicular direction. The other points of the string are moving towards the minimum. 

\textit{Step 2:} To ensure that the endpoint converges to an adjacent saddle point, one checks after $k$ timesteps whether the energy rises monotonically along the string. If there is an image $\boldsymbol{\Phi}_J$ with $H(\boldsymbol{\Phi}_J) > H(\boldsymbol{\Phi}_{J+1})$, the string is truncated at $\boldsymbol{\Phi}_J$. Furthermore, the set of images is rebalanced such that they remain evenly distributed along the string. We parametrize the  images in terms of the arc\-length,  computed as 
\begin{equation}
s_0 = 0, \ \ s_{i+1} = s_i + |\boldsymbol{\Phi}_{i+1} - \boldsymbol{\Phi}_{i}|.
\end{equation}
The parameters of the images are then given by
\begin{equation}
\alpha^*_i = s_i/s_J.
\end{equation}
The images with their associated parameters are used to calculate an analytic expression for the string via, e.g., cubic interpolation. The curve $\gamma = \{ \boldsymbol{\Phi}(\alpha),\ \  \alpha \in [0,1]\}$ is evaluated at $\alpha = i/M$ for $0 \leq i \leq M$ to recover a set of evenly distributed images. 

\textit{Step 3:}  We check whether the endpoint of the string converged to a saddle point by the condition
\begin{equation}
\max\{|\boldsymbol{\nabla} H(\boldsymbol{\Phi}_M)|, \ \ 0 \leq i \leq N^2\} < \delta
\end{equation}
with a tolerance $\delta$. If not, one repeats the previous steps until convergence.  

For a striped vortex lattice configuration $\mathbf{\Phi_0} = (\phi_{11},\phi_{12},...)$ with a domain wall at site $({ij})$, we use the initial configuration $\mathbf{\Phi_M} = (\phi'_{11},\phi'_{12},...)$ with 
\begin{equation}
    \phi'_{i+1,j+1} = \phi_{i+1,j+1} + 2.
\end{equation}
and otherwise identical phases. This choice ensures that the saddle point is connected to a domain-wall hop. We further use $M=20$ images, a time step $\Delta t = 0.01$, 
a force parameter $\theta = 1.2$, and a tolerance $\delta=0.0001$. We rebalance the images after $k=500$ time steps. The bond energy $E_J$ is set to unity. 
 
\end{document}